\documentclass{ws-mplb}

\usepackage{latexsym}
\usepackage{revsymb}
\usepackage{amsfonts}
\usepackage{amsmath}
\usepackage{amssymb}
\usepackage{bm}

\newcommand{\vev}[1]{\langle #1\rangle}
\newcommand{\bra}[1]{\langle #1 |}
\newcommand{\ket}[1]{| #1 \rangle}
\newcommand{\braket}[2]{\langle #1 | #2\rangle}

\newcommand{\0}{\underline{0}}
\newcommand{\1}{\underline{1}}
\newcommand{\Fq}{\mathbb{Z}_q}

\begin{document}

\title{GALOIS FIELD QUANTUM MECHANICS}

\author{\footnotesize LAY NAM CHANG}
\address{Department of Physics, Virginia Tech, Blacksburg, VA 24061, USA\\
laynam@vt.edu}

\author{\footnotesize ZACHARY LEWIS}
\address{Department of Physics, Virginia Tech, Blacksburg, VA 24061, USA\\
zlewis@vt.edu}

\author{\footnotesize DJORDJE MINIC}
\address{Department of Physics, Virginia Tech, Blacksburg, VA 24061, USA\\
dminic@vt.edu}

\author{\footnotesize TATSU TAKEUCHI}
\address{Department of Physics, Virginia Tech, Blacksburg, VA 24061, USA\\
takeuchi@vt.edu}

\maketitle

\begin{abstract}
We construct a discrete quantum mechanics using a vector space over the Galois field $GF(q)$.
We find that the correlations in our model do not violate the
Clauser-Horne-Shimony-Holt (CHSH) version of Bell's inequality,
despite the fact that the predictions of this discrete quantum mechanics cannot be reproduced with any hidden variable theory.
\keywords{Quantum mechanics; Galois field; Bell's inequality; Clauser-Horne-Shimony-Holt bound.}
\end{abstract}

\ccode{03.65.Aa,03.65.Ta,03.65.Ud}

\section{Introduction}

Though it is almost a century since the inception of quantum mechanics (QM), its foundations
and origin remain quite puzzling and continue to inspire intense inquiry.
In this letter, we attempt to illuminate the connection between the mathematical structure of 
QM and its physical characteristics by constructing a `mutant' quantum mechanical model which shares many, 
but not all of the mathematical features of canonical QM.  
By investigating which characteristics of canonical QM survive the `mutation'
and which ones do not, we hope to clarify the relation between the mathematical genotype and the physical phenotype.

In canonical QM, the states of an $N$-level quantum system are described by vectors
in the Hilbert space $\mathcal{H}_\mathbb{C}=\mathbb{C}^N$.
In the following,
we introduce a `mutation' by replacing $\mathcal{H}_\mathbb{C}$ with 
$\mathcal{H}_q=\Fq^N$ \cite{MQT,Finkelstein:1996fn,Finkelstein:1983,nambu},
where $\Fq$ is shorthand for the finite Galois field $GF(q)$,
$q=p^n$ for some prime $p$, and $n\in\mathbb{N}$.
For the case $n=1$, we have $GF(p)=\mathbb{Z}/p\mathbb{Z}$.
Such replacements of the vector space have been considered previously, 
\textit{e.g.} real QM in which $\mathcal{H}_\mathbb{C}$ is replaced by $\mathcal{H}_\mathbb{R}=\mathbb{R}^N$ \cite{Stueckelberg:1960},
and quaternionic QM in which it is replaced by $\mathcal{H}_\mathbb{H}=\mathbb{H}^N$ \cite{Adler:1995}. 
However,
the vector space $\mathcal{H}_q$, in contradistinction to $\mathcal{H}_\mathbb{R}$, $\mathcal{H}_\mathbb{C}$,
or $\mathcal{H}_\mathbb{H}$, 
lacks an inner product, normalizable states, and symmetric/hermitian operators.
Nevertheless, we find that we can construct a perfectly `quantum' model on it,
which predicts probabilities of physical measurements that cannot be
reproduced in any hidden variable theory.
What will not survive this `mutation,' however, are the
super-classical correlations of canonical QM.
In particular, we show that in our discrete QM, 
the Clauser-Horne-Shimony-Holt (CHSH) 
\cite{Clauser:1969ny} version of 
Bell's inequality \cite{bell,bell2} is not violated.

Before we proceed, we emphasize that our model is distinct from 
`Galois quantum systems' discussed in the literature \cite{Vourdas:2004,Vourdas:2007}.
There, it is the phase space which is assumed to be $\Fq\times \Fq$,
that is, the position and momentum of a particle take on values in $\Fq$.
In our approach, it is the wave-functions that take on values in $\Fq^N$,
while the outcomes of measurements take on values in $\mathbb{R}$.

\section{The Model}

Our starting point is the following canonical expression for the
probability of obtaining the outcome represented by the dual-vector $\bra{x}\in\mathcal{H}_\mathbb{C}^*$
when a measurement is performed on the state represented by the 
vector $\ket{\psi}\in\mathcal{H}_\mathbb{C}$: 
\begin{equation}
P(x|\psi) \;=\; \dfrac{\bigl|\braket{x}{\psi}\bigr|^2}{\sum_y \bigl|\braket{y}{\psi}\bigr|^2}\;.
\label{Pdef}
\end{equation}
Here, $\ket{\psi}$ is not normalized and
the sum in the denominator runs over the duals of all the eigenstates of a hermitian operator
which represents the observable in question.
However, for this expression to be interpretable as a probability,
the necessary condition is that the dual-vectors in the sum
span the entire dual vector space $\mathcal{H}_\mathbb{C}^*$,
and any reference to operators acting on $\mathcal{H}_\mathbb{C}$ is inessential.
The interpretation that the bracket $\braket{x}{\psi}\in\mathbb{C}$ is
an inner product between two vectors also need not be imposed.
The probability depends only on the absolute values of the brackets
$|\braket{x}{\psi}|\in\mathbb{R}$.
Since we can multiply $\ket{\psi}$ with any non-zero complex number
without changing the probabilities defined via Eq.~(\ref{Pdef}),
we are compelled to identify
vectors which differ by a non-zero multiplicative constant
as representing the same physical state, endowing the
state space with the complex projective geometry
\begin{equation}
\mathbb{C}P^{N-1}
\;=\; (\,\mathbb{C}^{N}\backslash\{\mathbf{0}\}\,)\,\big/\,(\,\mathbb{C}\backslash\{0\}\,)
\;\cong\; S^{2N-1}\!\big/\,S^1\;,
\end{equation}
where each line going through the origin of $\mathbb{C}^N$ is identified as a `point.'

Thus, to construct a `mutant' QM on $\mathcal{H}_q$, 
we represent states with vectors $\ket{\psi}\in\mathcal{H}_q$, 
and outcomes of measurements with dual-vectors $\bra{x}\in\mathcal{H}_q^*$.
Observables are associated with a choice of basis of $\mathcal{H}_q^*$,
each dual-vector in it representing a different outcome.
The bracket $\braket{x}{\psi}\in\Fq$ is converted into
a non-negative real number $|\braket{x}{\psi}|\in\mathbb{R}$ 
via the absolute value function:
\begin{equation}
|\,\underline{k}\,|\;=\;
\left\{\begin{array}{ll}
0\quad &\mbox{if $\underline{k}=\0$}\;,\\
1\quad &\mbox{if $\underline{k}\neq\0$}\;.
\end{array}
\right.
\label{abs}
\end{equation}
Here, underlined numbers and symbols represent elements of $\Fq$, to
distinguish them from elements of $\mathbb{R}$ or $\mathbb{C}$.
Note that Eq.~(\ref{abs}) is not to be interpreted as a condition imposed on $\braket{x}{\psi}\in\Fq$;
all non-zero values of $\Fq$ are mapped to one.
Since $\Fq\backslash\{\0\}$ is a cyclic multiplicative group, 
this assignment of `absolute values' is the only one consistent with 
the requirement that the map from $\Fq$ to non-negative $\mathbb{R}$ be product preserving,
that is: $|\underline{k}\underline{l}|=|\underline{k}||\underline{l}|$.
With these assignments, Eq.~(\ref{Pdef}) can be applied as it stands to calculate probabilities.
Since the same absolute value is assigned to all non-zero brackets,
all outcomes $\bra{x}$ for which the bracket with the state $\ket{\psi}$
is non-zero are given equal probabilistic weight.

The product preserving nature of the absolute value function 
guarantees that the probabilities of product observables on product states
factorize in multi-particle systems:
\begin{eqnarray}
P(xy|\psi\phi)
& = & 
\dfrac{\bigl|\left(\bra{x}\otimes\bra{y}\right)\left(\ket{\psi}\otimes\ket{\phi}\right)\bigr|^2}
      {\sum_{zw}\bigl|\left(\bra{z}\otimes\bra{w}\right)\left(\ket{\psi}\otimes\ket{\phi}\right)\bigr|^2}
\cr
& = &
\dfrac{\bigl|\braket{x}{\psi}\braket{y}{\phi}\bigr|^2}
      {\sum_{zw}\bigl|\braket{z}{\psi}\braket{w}{\phi}\bigr|^2}
\;=\;
\dfrac{\bigl|\braket{x}{\psi}\bigr|^2\bigl|\braket{y}{\phi}\bigr|^2}
      {\sum_{zw}\bigl|\braket{z}{\psi}\bigr|^2\bigl|\braket{w}{\phi}\bigr|^2}
\cr
& = &
\dfrac{\bigl|\braket{x}{\psi}\bigr|^2}
      {\sum_{z}\bigl|\braket{z}{\psi}\bigr|^2}
\;
\dfrac{\bigl|\braket{y}{\phi}\bigr|^2}
      {\sum_{w}\bigl|\braket{w}{\phi}\bigr|^2}
\;=\;
P(x|\psi)\,P(y|\phi)\;.
\end{eqnarray}
This property is crucial if we want to have isolated particle states,
and is of course shared by canonical QM defined on $\mathcal{H}_{\mathbb{C}}$.

Note also that the multiplication of $\ket{\psi}$ with a non-zero element of
$\Fq$ will not affect the probability. 
Thus,
vectors that differ by non-zero multiplicative constants are identified as representing the same 
physical state, and the state space is endowed with the finite projective geometry
\cite{Hirschfeld,Hirschfeld2,Arnold,Ball-Weiner}
\begin{equation}
PG(N-1,q) \;=\; (\,\Fq^N\backslash\{\mathbf{\0}\}\,)\,\big/\,(\,\Fq\backslash\{\0\}\,)\;,
\end{equation}
where each `line' going through the origin of $\Fq^N$ is identified as a `point,'
in close analogy to the complex projective geometry of canonical QM.

\section{An Example}

To give a concrete example of our proposal, let us construct a 2-level system,
 analogous to spin, for which $\mathcal{H}_q = \Fq^2$, and the state space is
$PG(1,q)$.
This geometry consists of $q+1$ `points,'
which can be represented by the vectors
\begin{equation}
\ket{\,0\,} = \left[\begin{array}{c} \1 \\ \0 \end{array}\right],\quad
\ket{\,1\,} = \left[\begin{array}{c} \0 \\ \1 \end{array}\right],\quad
\ket{\,r\,} = \left[\begin{array}{l} \underline{a}^{r-1} \\ \1 \end{array}\right],
\end{equation}
$r=2,3,\cdots,q$, where $\underline{a}$ is the generator of the multiplicative group $\Fq\backslash\{\0\}$
with $\underline{a}^{q-1}=1$.
The number $q+1$ results from the fact that of the $q^2-1$ non-zero vectors, every $q-1$ are equivalent,
thus the number of inequivalent vectors are $(q^2-1)/(q-1)=(q+1)$.
Similarly, the $q+1$ inequivalent dual-vectors can be represented as:
\begin{eqnarray}
\bra{\,\overline{0}\,} & = & \bigl[\,\0\;-\!\1\,\bigr]\;,\cr
\bra{\,\overline{1}\,} & = & \bigl[\,\1\;\;\phantom{-}\0\,\bigr]\;,\cr
\bra{\,\overline{r}\,} & = & \bigl[\,\1\;-\!\underline{a}^{r-1}\,\bigr]\;,\qquad r=2,3,\cdots,q\;,
\end{eqnarray}
where the minus signs are dropped when the characteristic of $\Fq$ is two.
From these definitions, we find:
\begin{eqnarray}
\braket{\bar{r}}{s} 
& =    & \0\quad \mbox{if $r=s$}\;, \cr 
& \neq & \0\quad \mbox{if $r\neq s$}\;,
\end{eqnarray}
and
\begin{equation}
\bigl|\braket{\bar{r}}{s}\bigr|\;=\; 1-\delta_{rs}\;.
\label{braketpq}
\end{equation}
Observables are associated with a choice of basis of $\mathcal{H}_q^*$:
\begin{equation}
A_{rs}\;\equiv\;\{\;\bra{\bar{r}},\;\bra{\bar{s}}\;\}\;,\qquad
r\neq s\;.
\end{equation}
We assign the outcome $+1$ to the first dual-vector of the pair, 
and the outcome $-1$ to the second to make these observables
spin-like. This assignment implies $A_{sr}=-A_{rs}$.
The indices $rs$ can be considered as indicating the direction of the `spin,'
and the interchange of the indices as indicating a reversal of this direction.

Applying Eq.~(\ref{Pdef}) to this system, it is straightforward to show that
\begin{eqnarray}
P(A_{rs}=+1\,|\,r) & = & 0\;,\qquad
P(A_{rs}=-1\,|\,r) \;=\; 1\;,\cr
P(A_{rs}=+1\,|\,s) & = & 1\;,\qquad
P(A_{rs}=-1\,|\,s) \;=\; 0\;,\cr
P(A_{rs}=\pm 1\,|\,t) & = & \frac{1}{2}\;,\qquad\mbox{for $t\neq r, s$}\;,
\end{eqnarray}
and thus,
\begin{eqnarray}
\vev{A_{rs}}_r & = & -1\;,\cr
\vev{A_{rs}}_s & = & +1\;,\cr
\vev{A_{rs}}_t & = & \phantom{-}0\;,\quad\mbox{for $t\neq r, s$.}
\end{eqnarray}
So for each `spin,' there exist two `eigenstates,'
one for $+1$ (`spin' up) and another for $-1$ (`spin' down).
For all other states the two outcomes $\pm 1$ are equally probable.

The states and observables `rotate' into each other under changes of bases.
For the projective geometry $PG(1,q)$, the group of all possible basis transformations
constitute the projective group $PGL(2,q)$ of order $q(q^2-1)$.
$PGL(2,q)$ is formally a subgroup of $S_{q+1}$, the group of all possible permutations of the $q+1$ states.

\section{Spin Correlations}

To show that our system is truly quantum, we use an argument 
analogous to those of Greenberger, Horne, Shimony, and Zeilinger \cite{GHZ,GHSZ},
and of Hardy \cite{Hardy:1993zza} for canonical QM.
Let us construct
a two `spin' system on the tensor product space $\Fq^2\otimes\Fq^2 = \Fq^4$.
The number of non-zero vectors in this space is
$q^4-1$, of which every $q-1$ are equivalent, so the number of inequivalent
states is $(q^4-1)/(q-1)=q^3+q^2+q+1$.
Of these, $(q+1)^2$ are product states, leaving
$(q^3+q^2+q+1)-(q+1)^2=q(q^2-1)$ that are entangled.
As noted previously, Eq.~(\ref{Pdef}) applied to tensored spaces
with the product preserving absolute value function Eq.~(\ref{abs}) 
ensures that the expectation values of product observables
factorize for product states, thereby rendering the distinction between
product and entangled states meaningful.

The number of entangled states matches the order of the group $PGL(2,q)$,
since arranging the 4 elements of an entangled state into a $2\times 2$ array
gives rise to a non-singular matrix.
The entangled states fall into `conjugacy' classes, matching those of $PGL(2,q)$,
that transform among themselves under $PGL(2,q)$ `rotations.'
The singlet state, corresponding to the conjugacy class of the unit element,
can be expressed as
\begin{equation}
\ket{S} \;=\; \ket{r}\otimes\ket{s}-\ket{s}\otimes\ket{r}\;,\qquad r\neq s\;,
\end{equation}
for any two states $\ket{r}$ and $\ket{s}$ up to a multiplicative constant.
If the characteristic of $\Fq$ is two, the minus sign is replaced by a plus sign.

Products of the `spin' observables are defined as
\begin{equation}
A_{rs}A_{tu}
\,=\,\{
\,\bra{\bar{r}}\otimes\bra{\bar{t}}\,,
\,\bra{\bar{r}}\otimes\bra{\bar{u}}\,,
\,\bra{\bar{s}}\otimes\bra{\bar{t}}\,,
\,\bra{\bar{s}}\otimes\bra{\bar{u}}\,
\}\;,
\end{equation}
the four tensor products representing the outcomes
$++$, $+-$, $-+$, and $--$,
and the expectation value giving the correlation between the two `spins.' 
The probabilities of the four outcomes are
particularly easy to calculate for the singlet state $\ket{S}$ since \cite{MQT}
\begin{eqnarray}
\bigl(\bra{\bar{r}}\otimes\bra{\bar{s}}\,\bigr)\ket{S}
& =    & \0\quad \mbox{if $r=s$}\;, \cr
& \neq & \0\quad \mbox{if $r\neq s$}\;,
\end{eqnarray}
thus
\begin{equation}
\Bigl|
\bigl(\bra{\bar{r}}\otimes\bra{\bar{s}}\,\bigr)\ket{S}
\Bigr|
\;=\; 1-\delta_{rs}\;,
\end{equation}
and we obtain the probabilities and correlations listed in Table.~\ref{Probs}.

\begin{table}
\begin{center}
\begin{tabular}{|c||c|c|c|c||c|}
\hline
\ Observable\ \ &\ $++$\ \ &\ $+-$\ \ &\ $-+$\ \ &\ $--$\ \ &\ E.V. \ \\
\hline
$A_{rs}A_{rs}$ & $0$            & $\dfrac{1}{2}$ & $\dfrac{1}{2}$ & $0$            & $-1$ \phantom{\bigg|} \\
\hline
$A_{rs}A_{rt}$ & $0$            & $\dfrac{1}{3}$ & $\dfrac{1}{3}$ & $\dfrac{1}{3}$ & $-\dfrac{1}{3}$ \phantom{\bigg|}\\
\hline
$A_{rs}A_{st}$ & $\dfrac{1}{3}$ & $\dfrac{1}{3}$ & $0$            & $\dfrac{1}{3}$ & $+\dfrac{1}{3}$ \phantom{\bigg|}\\
\hline
$A_{rs}A_{tu}$ & $\dfrac{1}{4}$ & $\dfrac{1}{4}$ & $\dfrac{1}{4}$ & $\dfrac{1}{4}$ & $\phantom{-}0$  \phantom{\bigg|}\\
\hline
\end{tabular}
\caption{Probabilities and expectation values of product observables in the singlet state $\ket{S}$.
The indices $r$, $s$, $t$, and $u$ are distinct.
Cases that can be obtained by flipping signs using $A_{rs}=-A_{sr}$ are not shown.}
\label{Probs}
\end{center}
\end{table}

To demonstrate that these correlations cannot be reproduced in 
any hidden variable theory, it suffices to look at the correlations between two observables that share an index.
For instance, consider the following two:
\begin{equation}
X\;\equiv\;A_{01}\;,\quad
Y\;\equiv\;A_{02}\;.
\end{equation}
First, from the first row of Table~\ref{Probs} we can discern that 
\begin{equation}
\begin{array}{lll}
P(X_1X_2;++|S) & =\; P(X_1X_2;--|S) & =\; 0\;,\\
P(Y_1Y_2;++|S) & =\; P(Y_1Y_2;--|S) & =\; 0\;,
\end{array}
\end{equation}
where we have added subscripts to distinguish between the two `spins.'
This tells us that the pairs $(X_1X_2)$
and $(Y_1Y_2)$ are completely anti-correlated. 
Next, from the second row of Table~\ref{Probs}, we conclude:
\begin{equation}
P(X_1Y_2;++|S) \;=\; P(Y_1X_2;++|S) \;=\; 0\;,
\end{equation}
which means that if either one of the pairs $(X_1Y_2)$ and $(Y_1X_2)$ is
$+1$, then its partner must be $-1$.
Thus, the implications of either $X_1=+1$ or $X_1=-1$ would be:
\begin{equation}
\begin{array}{llll}
X_1=+1 &\rightarrow\;
Y_2=-1 &\rightarrow\;
Y_1=+1 &\rightarrow\;
X_2=-1 \;,\\
X_1=-1 &\rightarrow\;
X_2=+1 &\rightarrow\;
Y_1=-1 &\rightarrow\;
Y_2=+1 \;.\\
\end{array}
\end{equation}
In either case, we cannot classically have $(X_1Y_2)=(--)$ 
or $(Y_1X_2)=(--)$, even though both configurations have quantum mechanical
probabilities of $1/3$.
Thus, our `mutant' QM is truly `quantum' and its predictions do not
allow any hidden variable mimic.

Let us now look at what the Clauser-Horne-Shimony-Holt (CHSH) bound \cite{Clauser:1969ny}
would be in our `mutant' QM.  The CHSH bound is
the upper bound of the absolute value of the following combination of correlators:
\begin{equation}
\vev{A,a\,;B,b}
\;\equiv\;\vev{AB}+\vev{Ab}+\vev{aB}-\vev{ab}
\;,
\label{CHSHcorr}
\end{equation}
where $A$ and $a$ are two observables of particle 1, and
$B$ and $b$ are two observables of particle 2.
All four observables are assumed to take on only the values $\pm 1$ upon 
measurement.
For classical hidden variable theory, the bound
on $\left|\vev{A,a;B,b}\right|$ is 2, while for
canonical QM it is $2\sqrt{2}$ \cite{cirelson,landau}.

To calculate this bound for our model, it suffices to examine all possible correlators
for the singlet state $\ket{S}$ only.
This is because all $q(q^2-1)$ entangled states can be transformed
into $\ket{S}$ via local $PGL(2,q)$ rotations, that is, $PGL(2,q)$ transformations on only one of 
the entangled particles.
We can also restrict the observables entering the correlator to those in which the indices
are in increasing order, \textit{i.e.} $A_{rs}$ with $r<s$, since
\begin{eqnarray}
\vev{A,a\,;B,b}
& = & \vev{A,-a\,;b,B}
\;=\; -\vev{-A,a\,;b,B} \cr
& = & \vev{a,A\,;B,-b}
\;=\; -\vev{a,A\,;-B,b}\;.\quad
\end{eqnarray}
These considerations simplify our task considerably, and
we find that the absolute value of the CHSH correlator is maximized for
the cases
\begin{eqnarray}
\vev{A_{rs},A_{tu};A_{tu},A_{rs}}_S & = & -2\qquad (r<s,\;t<u)\;,\cr
\vev{A_{rs},A_{st};A_{rt},A_{rs}}_S & = & -2\qquad (r<s<t)\;.
\end{eqnarray}
Thus, the CHSH bound for our `mutant' QM is 2.

\section{Conclusion}

In conclusion, we have constructed a `mutant' QM based on a linear
vector space over the Galois field $\Fq=GF(q)$.
We find that though it is fully `quantum' in the sense that no hidden variable theory
can reproduce its predictions, the CHSH bound of its correlations nevertheless
has the `classical' value of 2.
Thus, our model provides an existence proof that `quantum'-ness does
not necessarily require the violation of the CHSH bound.

The state space of our `mutant' QM is the finite projective space $PG(N-1,q)$, 
in close analogy to the $\mathbb{C}P^{N-1}$ of canonical QM.
We recall that this complex projective space can be understood as the coset 
\begin{equation}
\mathbb{C}P^{N-1} \;\cong\; U(N)\,\big/\,\bigl(\,U(N-1) \times U(1)\,\bigr)\;.
\end{equation}
This structure incorporates unitary evolution described by the $U(N)$ factor, 
thereby preserving the normalization of state vectors, 
with generic Berry phases described by the $U(N-1)$ factor, 
which characterize possible degenerate states, 
and the quantum mechanical $U(1)$ phases \cite{Ashtekar:1997ud}.
The corresponding coset structure of $PG(N-1,q)$ is :
\begin{equation}
PG(N-1,q) \;\cong\; GL(N,q)\,\big/\,\bigl(AGL(N-1,q)\times Z(N,q)\,\bigr)\;,
\end{equation}
where, $GL(N,q)$ is the general linear group on $\mathcal{H}_q$,
$Z(N,q)$ its center consisting of $N\times N$ unit matrices multiplied by a `phase' in $\Fq\backslash\{\0\}$,
and $AGL(N-1,q)$ is the affine linear group which keeps the direction of a vector in $\mathcal{H}_q$ invariant.
The projective linear group we encountered earlier is itself the coset group
\begin{equation}
PGL(N,q) \;=\; GL(N,q)\big/Z(N,q)\;.
\end{equation}
Thus, our discrete QM possesses analogs of the geometric structure of canonical QM, 
with $GL(N,q)$ generating evolution over finite time steps, 
and $AGL(N-1,q)$ characterizing possible degeneracies in dynamical systems.   
The extent that elements of this affine group and the center $Z(N,q)$ 
determines any super-selection sectors has yet to be explored \cite{Jejjala:2007rn}.

The $q=2$ case of our model, constructed on $\mathbb{Z}_2=\{\0,\1\}$, would be particularly simple.
It could, perhaps, be the simplest quantum theory imaginable 
and provide a setting to explore the most basic questions concerning the foundations of QM,
as well as a platform to develop ideas relevant to quantum information and quantum computing
\cite{James:2011}.

A question we addressed in a previous publication \cite{Chang:2011yt} was
whether a super-quantum theory whose CHSH bound exceeds the  
Cirel'son value of $2\sqrt{2}$ of canonical QM exists \cite{cirelson,super}.
Such super-quantum models are expected to go `beyond' canonical QM in one way 
or another.
The model discussed in this paper was discovered in the process of 
looking for such a model, though, of course, its correlations are sub-quantum instead.
An interesting question to ask is whether a `super' version of our 
discrete model can be constructed in which the CHSH bound exceeds 2.
We conjectured in Ref.~\refcite{Chang:2011yt} that a super-quantum theory may exist 
in the $\hbar\rightarrow\infty$ limit of canonical QM.
Recalling that $1/\hbar$ is effectively the curvature of $\mathbb{C}P^{N-1}$, 
this complex projective space would degenerate to $\mathbb{C}^{N-1}$ in the $\hbar\rightarrow\infty$ limit.
Extrapolating this intuition to our discrete QM, whose geometry is described by
$PG(N-1,q)=(\Fq^N\backslash\{\mathbf{\underline{0}}\})/(\Fq\backslash\{\0\})$, 
the super-quantum limit would correspond to degenerating this
projective space into $\Fq^{N-1}$. 
Thus, the construction of a `super' version of discrete QM may require
working in this space.
 
Another interesting avenue, already alluded to above, 
would be to use our discrete QM to explore the structure of 
the conjectured general geometric quantum theory \cite{Jejjala:2007rn}.
Such a structure was argued to be relevant for quantum gravity.
In particular, in that context, it was argued that non-linear Grassmannian spaces are the natural
generalization of complex projective spaces \cite{Minic:2003en,Minic:2003nx,Minic:2004rj}.  
It view of our proposal, it would be interesting to explore
discrete analogs of non-linear Grassmannians by replacing complex spaces with $\Fq$-spaces.

Finally, since we have succeeded in constructing a QM on a space without an inner product, 
it cannot be an essential gene necessary for the survival of a `quantum' theory. 
The absence of an inner product allows the vector and dual-vector spaces to be distinct, 
and our construction gives separate physical meanings to the two: the dual-vectors represent possible outcomes
of a measurement, while the vectors represent the latent possibilities of a state. 
This removal of the required existence of an inner product potentially allows 
for the construction of quantum theories on spaces that have not heretofore been considered,
\textit{e.g.} Banach spaces.

We will explore these, and other related questions in upcoming publications \cite{next}.

\section*{Acknowledgments}

We would like to thank Sir Anthony Leggett and Prof. Chia Tze for helpful discussions.  
ZL, DM, and TT are supported in part by
the U.S. Department of Energy, grant DE-FG05-92ER40677, task A.



\begin{thebibliography}{99}


\bibitem{MQT}
 A similar proposal was made by	
 B.~Schumacher and M.~D.~Westmoreland in
 arXiv:1010.2929 [quant-ph].
 In their work, probabilities were not defined.
 Our model would correspond to assigning equal probabilities to
 all `possible effects' in their model.


\bibitem{Finkelstein:1996fn}
  D.~R.~Finkelstein,
  ``Quantum relativity: A Synthesis of the ideas of Einstein and Heisenberg,''
  Springer, 1996, pages 76-78.


\bibitem{Finkelstein:1983} 
  D.~Finkelstein and S.~R.~Finkelstein,
  Int.\ J.\ Theor.\ Phys.\  {\bf 22} (1983) 753.


\bibitem{nambu} 
 For a pioneering discussion of quantum field theory over Galois fields consult:
 Y.~Nambu,
  ``Field Theory of Galois Fields,''
  in \textit{Quantum Field Theory and Quantum Statistics}, Vol.~1, pp. 625-636,
  eds. I.~A.~Batalin et al. (IOP Publishing, 1987).


\bibitem{Stueckelberg:1960} 
  E.~C.~G.~Stueckelberg,
  Helv.\ Phys.\ Acta {\bf 33} (1960) 727.

\bibitem{Adler:1995}
 S.~L.~Adler, {\it Quaternionic Quantum Mechanics and Quantum Fields},
 Oxford University Press, 1995.

\bibitem{Clauser:1969ny}
  J.~F.~Clauser, M.~A.~Horne, A.~Shimony and R.~A.~Holt,
  Phys.\ Rev.\ Lett.\  {\bf 23} (1969) 880.


\bibitem{bell}
 J.~S.~Bell, Physics {\bf 1} (1964) 195.


\bibitem{bell2}
 J.~S.~Bell, {\it Speakable and Unspeakable in Quantum Mechanics}, Cambridge University Press (1987).
 
\bibitem{Vourdas:2004}
  A.~Vourdas,
  Rep.\ Prog.\ Phys.\ {\bf 67} (2004) 267.


\bibitem{Vourdas:2007}
  A.~Vourdas,
  J.\ Phys.\ A\ {\bf 40} (2007) R285.



\bibitem{Hirschfeld}
 J.~W.~P.~Hirschfeld, {\it Projective Geometries over Finite Fields,} 2nd ed.,
 Oxford University Press, 1998.


\bibitem{Hirschfeld2}
 J.~W.~P.~Hirschfeld, G.~Korchm\'aros, and F.~Torres,
 {\it Algebraic Curves over a Finite Field},
 Princeton University Press, 2008.


\bibitem{Arnold}
 V.~I.~Arnold, {\it Dynamics, Statistics and Projective Geometry of Galois Fields},
 Cambridge University Press, 2011.

\bibitem{Ball-Weiner}
 S.~Ball and Z.~Weiner, {\it An Introduction to Finite Geometry,}
 \texttt{http://www-ma4.upc.es/\~{}simeon/IFG.pdf}

\bibitem{GHZ}
  D.~M.~Greenberger, M.~A.~Horne, A.~Zeilinger,
  arXiv:0712.0921v1 [quant-ph].
  
\bibitem{GHSZ}  
  D.~M.~Greenberger, M.~A.~Horne, A.~Shimony, and A.~Zeilinger,
  Am.\ J.\ Phys.\ {\bf 58} (1990) 1131.

\bibitem{Hardy:1993zza}
  L.~Hardy,
  Phys.\ Rev.\ Lett.\  {\bf 71} (1993) 1665.


\bibitem{cirelson}
 B. S. Cirel'son, Lett. Math. Phys. {\bf 4} (1980) 93.
 

\bibitem{landau}
 L. J. Landau, Phys. Lett. {\bf A 120} (1987) 52.


\bibitem{Ashtekar:1997ud} 
  A.~Ashtekar and T.~A.~Schilling,
  gr-qc/9706069.


\bibitem{Jejjala:2007rn} 
  For a review see: V.~Jejjala, M.~Kavic and D.~Minic,
  Int.\ J.\ Mod.\ Phys.\ A {\bf 22} (2007) 3317.



\bibitem{James:2011}
 See, for instance:
 R.~P.~James, G.~Ortiz, and A.~Sabry,
 arXiv:1101.3764 [quant-ph].



\bibitem{Chang:2011yt}
  L.~N.~Chang, Z.~Lewis, D.~Minic, T.~Takeuchi, and C.~H.~Tze,
  Advances in High Energy Physics {\bf 2011} (2011) 593423 .

\bibitem{super}
 S.~Popescu and D.~Rohrlich, Foundations of Physics, {\bf 24} (1994) 379.


\bibitem{Minic:2003en} 
  D.~Minic and C.~H.~Tze,
  Phys.\ Rev.\ D {\bf 68} (2003) 061501.


\bibitem{Minic:2003nx} 
  D.~Minic and C.~H.~Tze,
  Phys.\ Lett.\ B {\bf 581} (2004) 111.


\bibitem{Minic:2004rj} 
  D.~Minic and C.~H.~Tze,
  hep-th/0401028.



\bibitem{next}
 L.~N.~Chang, Z.~Lewis, D.~Minic, and T.~Takeuchi,
 arXiv:1206.0064,
 arXiv:1208.5189,
 arXiv:1208.5544,
 and other papers in preparation.

 


\end{thebibliography}
\end{document}